# High-speed laser writing of structural colors for full-color inkless printing

Jiao Geng[1,2], Liye Xu[1,2], Wei Yan[1,2], Liping Shi[1,2*] and Min Qiu[1,2*]

[1]Key Laboratory of 3D Micro/Nano Fabrication and Characterization of Zhejiang Province, School of Engineering, Westlake University, 18 Shilongshan Road, Hangzhou 310024, Zhejiang Province, China.

[2]Institute of Advanced Technology, Westlake Institute for Advanced Study, 18 Shilongshan Road, Hangzhou 310024, Zhejiang Province, China.

*Corresponding author(s). E-mail(s): shiliping@westlake.edu.cn; qiumin@westlake.edu.cn;

**Abstract**

It is a formidable challenge to simultaneously achieve wide gamut, high resolution, high-speed while low-cost manufacturability, long-term stability, and viewing-angle independence in structural colors for practical applications. The conventional nanofabrication techniques fail to match the requirement in low-cost, large-scale and flexible manufacturing. Processing by ultrashort lasers can achieve extremely high throughput while suffering from a narrow gamut of ∼15% sRGB or angle-dependent colors. Here, we demonstrate an all-in-one solution for ultrafast laser-produced structural colors on ultrathin hybrid films that comprise an absorbent TiAlN layer coating on a metallic TiN layer. Under pulsed laser irradiation, the absorption behaviors of the TiAlN-TiN hybrid films are tailored by photothermal-induced oxidation on the topmost TiAlN. The oxidized films exhibit double-resonance absorption, which is attributed to the non-trivial phase shifts both at the oxide-TiAlN interface, and at the TiAlN-TiN interface. By varying the accumulated laser fluence to modulate the oxidation depth, an unprecedented large gamut of ∼90% sRGB is obtained. Our highly reproducible printing technique manifests angle-insensitive colors the variation of Hue is < 0.14π when viewing angles changing from 6 to 60°. The full-color printing speed reaches to 1.4 cm$^2$/s and the highest printing resolution exceeds 25000 dpi. The durability of the laser- printed colors is confirmed by fastness examination, including salt fog, double-85, light bleaching, and adhesion tests. These features render our technique to be competitive for high-throughput industrial applications.

# 1 Introduction

Inkjet or laser printers rely on ink or toner cartridges, in which colorful pigments serve to selectively absorb visible light within a spectral range. However, the conventional pigments are toxic and environment unfriendly. Further, their long-term stability in ambient is generally poor, because the organic pigments tend to degrade over time and thus losing their chroma as well as brightness. [1] Therefore, the development of structural colors-based ink-free printers is demanded on looming. In recent years, the structural colors [2]— arising from light scattering, absorption, diffraction or interference by micro-/nanostructures including plasmonic [3–10] and all-dielectric metasurfaces [11–19], diffractive elements, [20–23] micro-fibrillation, [24] Fabry-Perot (FP) cavities, [25–29] multilayered dielectric films (Bragg mirror), [30, 31] Fano-resonance optical coatings, [32] etc — are promising to achieve pigments- free colorful printing. It is hence becoming an attractive technique in many applications, such as the labeling of serial number, barcode, quick response code, company logo, trade marker, anti-counterfeiting, to name a few.

However, surface coloring in terms of the conventional nanofabrication techniques, such as electron beam lithography, [33, 34] focused ion beam milling [35] and nanoimprinting lithography, [36] is facing the nanoscale and macroscale processing barrier. Production of large-scale colored surfaces with these techniques is incompatible with the demands of low-cost mass manufacturing. As an alternative, surface coloring by pulsed lasers is appealing to overcome this barrier. [37] Laser coloring generally contains three approaches: plasmonic colors from randomly self-organized metallic nanoparticles, [38] diffractive colors from laser-induced periodic nanogratings, [23, 39], and interfering colors from thin films, e.g. FP-cavities. [26, 40, 41]

Nevertheless, the plasmonic colors have low stability, narrow gamut (15% sRGB), [38] moderate productivity while high operation cost. [37] The FP-cavities consisting of metal-dielectric-metal (MDM) is obtained by laser- induced modification of the topmost metallic film. It covers relatively narrow gamut (20% sRGB) [26] and low image contrast. Further, both plasmonic and MDM-interfering colors rely on noble metals such as Au and Ag, exhibiting low wear resistance and thus poor abrasion stability. [26, 37, 40] The ripples produce iridescent colors via diffraction of nanogratings. The iridescence is undesired for many applications. [37]

So far, the widely adopted laser coloring technique is in terms of laser- induced oxidation, generally occurring on metallic surfaces such as titanium or stainless steel. [37, 42] This approach exhibits the advantages of high stability, productivity and low operation cost. The colors come from interference of reflection from top and bottom surfaces of the laser-oxidized transparent film. The optical path difference $\delta$ between the two reflected beams can be expressed by $\delta = 2hn_1\sqrt{n_1^2 - sin^2\theta_i}$, where $h$, $n1$ and $\theta_i$ denote the thickness and refractive index of oxide film, and the incidence angle, respectively. At normal incidence ($\theta_i$=0), when $\delta$ = $2hn_1$ = $\lambda$ (i.e., $h$ = $\lambda/2n_1$) for a wavelength $\lambda$, the specific color originates from constructive interference-enhanced reflection, the same effect that is responsible for soap bubbles. However, in such cases, the reflected colors are dependent on viewing angles. [37, 43] Additionally, its gamut is also not wide enough (<30% sRGB) [37, 44]. Red and green colors are difficult to be achieved. This has been attributed to the film thickness $h$, which requires large enough for the origin of these colors associated with long-wavelength. [44]

## 1  Mechanisms

Here, we demonstrate a new scheme to solve the aforementioned problems. Wide gamut, durable, reproducible and viewing angle-insensitive structural colors are simultaneously achieved. As illustrated in Fig. 1 (a), highly absorbing dielectric TiAlN, coating on a reflective TiN film, serves as 'inorganic ink' for laser color printing. Such bilayer films, acting as a broadband absorber, represent thickness-dependent colors, [32, 45, 46] as displayed by the photographs in Fig. 1(b, i-iii). However, its gamut is rather narrow, as shown in International Commission on Illumination (CIE) 1931 *x-y* chromaticity diagram (blue curve in Fig. 1c). But interestingly, when depositing an ultrathin trans- parent dielectric such as AlN or $Al_2O_3$ film onto the TiAlN-TiN absorber, the observed colors can be dramatically changed [Fig. 1(b, iv-vi)]. The corresponding gamut is obviously extended (red and black curves in Fig. 1c). Especially, red and green colors are acquired.

Changing structural colors of the thin-film absorber by a transparent dielectric coating can be understood from the Snell's law. [47] At incidence angle of $\theta_i$ and refraction angle of $\theta_r$, the reflection coefficient at an interface is

$$\tilde{r}_{m,n} = \frac{\tilde{n}_m \cos(\theta_i) - \tilde{n}_n \cos(\theta_r)}{\tilde{n}_m \cos(\theta_i) + \tilde{n}_n \cos(\theta_r)} \quad (1)$$

where $\tilde{n}_{m,n} = n_{m,n} + ik_{m,n}$ is the complex-valued refractive index of medium *m* and *n*, with {*m, n*} = {1,2} or {2,3}. Assuming laser-oxidized layer has a refractive index close to $Al_2O_3$ (medium 1, $h_1$ = 40 nm, $\tilde{n}_1$ = 1.7), the stacking of medium 1 onto medium 2 (TiAlN: $\tilde{n}_2$ = 2.6+0.95i) exists a resonance absorption at 330 nm in case of normal incidence ($\theta_{i,r}$ = 0°), as numerically simulated in Fig. 1(d). According to Eq. (1), the reflection at the interface between them has a non-trivial phase shift $\varphi_{1,2} = 0.2\pi$. The total phase shift is $\psi = \varphi_{1,2} + 4h_1 n_1 \pi/\lambda = \pi$. Therefore, the corresponding colors are caused by *destructive interference*-induced resonance absorption, differing from the mechanisms of soap bubbles as mentioned above. Likewise, the non-trivial phase shift between medium 2 and medium 3 also results in a resonance absorption at 600 nm ($\varphi_{2,3} = 0.4\pi$), as shown in Fig. 1(e). The joint behavior of these three layers leads to a double-resonance absorption (Fig. 1e), which is confirmed by experiments (Fig. 1f). Such a double resonance gives rise to more freedoms to tune the optical absorption and thus resulting in a wider gamut than the original TiAlN-TiN absorber. More importantly, as these coatings are much thinner than the visible light wavelength, the phase accumulation due to the propagation through the film is small. Therefore, they display viewing angles-insensitive colors.

## 3 Results

In experiments, the lossless dielectric layer is formed through laser-induced oxidation on the surface of TiAlN. As shown in Fig. 1(a), a linearly polarized incident picosecond laser is rotated by a quarter-wave plate to circular polarization (see Methods). This is to avoid the unwanted iridescence that comes from diffraction of laser-induced nanoripples. [23, 39, 48] Figure 2(a) shows a matrix palette that is produced on 60-nm-TiAlN coating on 50-nm-TiN by raster scanning. Various vibrant colors spanning from red, orange, yellow, green, blue to purple are observed. The applied laser repetition rate is f = 5 kHz and the laser spot focal waist diameter is $\sigma$ = 120 μm. The colors are dependent on the scanning speeds ($v_s$ ranging from 1 to 400 mm/s) and laser pulse energy (E spanning from 26 to 50 μJ). These parameters define the total accumulated laser fluence $F_t = \frac{2E}{\pi\sigma^2} N_{eff}$, with $N_{eff} = \frac{f\sigma}{v_s}$ denoting the effective number of irradiated pulses. Nevertheless, the much larger ranges of $v_s$ (or $N_{eff}$) with respect to that of E suggests that the oxidation process is more sensitive to the latter. This can be attributed to that the oxidation depth does not linearly increase with $N_{eff}$, because the penetration of $O_2$ through the formed oxide layer decreases exponentially, decelerating and eventually halting the growth process. The generated colors can be locally altered by a repeated scanning, which changes $F_t$, as shown in Fig. 2(b, c).

The repetition rate f and scanning speeds vs can be proportionally increased. When *f* = 2 MHz, vs can be increased up to 2 m/s. The printing resolution is essentially limited to the beam diameter but is smaller than it due to the threshold effect and the Gaussian distribution of laser fluence. The printing resolution is tested by producing colors with a single shot, which decides the ultimate limit of the pixel size. The diameter of a typical printing spot is 70 μm. Therefore, the spatial resolution of our setup is evaluated to be 2.54

cm/70 µm = 360 dpi, and the corresponding printing speed reaches to 2 m/s×70 µm = 1.4 cm$^2$/s. The resolution can be improved by utilizing a smaller laser beam spot. For instance, when $\sigma$ is reduced to the diffraction limit of 1 µm, the printing resolution will be greater than 20000 dpi. The spot sizes of pigments are generally on the order of 25 µm, resulting in a resolution below 1000 dpi. Therefore, the laser-printing ink-free structural colors can have much higher resolution than the conventional printers.

The reflectance spectra of several representative colors are plotted in Fig. 2(d). Indeed, we find that the laser-treated films exhibit double-resonance absorption, in agreement with the simulation in Fig. 1(e) and intentionally sputtered three-layer film (Fig. 1f). Plotting the reflectance spectra of the matrix palette into CIE 1931 chromaticity diagram, as shown in Fig. 2(e), we verify that the laser-printed colors have a wide gamut (~90% sRGB), which is obviously broader than that of the pristine TiAlN-TiN bilayer. The reflectance spectra versus incidence angles are plotted in Fig. 2(f). The spectral pro- files and the reflected peaks are nearly unchanged, suggesting that the colors are rather insensitive to the viewing angles. Plotting the spectra into a hue-saturation-lightness (HSV) diagram, we confirm that the colors are indeed robust against the viewing angles. Therefore, compared to laser-written colors directly on metals, [37] our technique exhibits the advantages of viewing angle-independence and wider gamut. Meanwhile, we remedy the drawbacks of the conventional nanofabrication methods such as low-throughput and high cost (Fig. 2g). [3]

In order to confirm that the surface coloring indeed arises from laser- induced oxidation, we perform extensive surface material analysis by energy dispersive x-ray spectroscopy (EDX), focused ion beam (FIB) milling, atomic force microscopy (AFM), x-ray diffraction (XRD), and x-ray photoelectron spectroscopy (XPS). Figure 3(a-e) list the SEM image of a laser-written area and corresponding two-dimensional EDX maps of oxygen, nitrogen, titanium and aluminum, respectively. The EDX maps confirm that, indeed, nitrogen has been partially replaced by oxygen, while the components titanium and aluminum are nearly unchanged. Next, the multilayered films are milled by a Ga+-FIB to observe their cross-sectional views, as depicted in Fig. 3(f-j). One can see that, for light blue color that is obtained at lowest laser power (inset in Fig. 3f), the oxidation degree is rather weak. When the color becomes to green (inset in Fig. 3g), three layers are clearly observed (Fig. 3g), which is in accordance with the numerical simulations in Fig. 1(e). For dark blue, pink and yellow colors that are obtained at higher accumulated fluence, the thickness of oxidation layer decreases with the increase of fluence. This indicates that the oxide layer has been partially ablated at high fluence. The surface roughness is measured to be Ra = 5 nm by

AFM (Fig. 3k,3l). The laser-modified mate- rial properties are further investigated by XRD. As shown in Fig. 3(m), the pristine sputtering TiAlN exhibit predominant peaks at 2θ=37.5° (black curve in Fig. 3m), representing a face centered cubic NaCl-type phase. Such phase is associated with high hardness, superior mechanical and tribological proper- ties. In addition, a small amount of $Al_2O_3$ at 2θ=43.5° that originates from natural oxidation is also detected. The XRD spectrum of green-colored area indicates the coexistence of $Al_2O_3$ and $TiO_2$ (green curve in Fig. 3l). With regard to the yellow-colored area, one can only observe $TiO_2$. This agrees with the cross-sectional observation in Fig. 3(j), which suggests that at high accumulated fluence, the TiAlN has been ablated and thus leaving TiN to be slightly oxidized. Further, when using $Ar^+$ to gradually etch the laser-irradiated area and investigate the depth profile of oxygen (Fig. 3n), we find that the amount of oxygen exponentially decreases with the increase of etching depth. This suggests that the oxidation depth does not linearly increase the accumulated number of pulses.

To illustrate the capability of printing arbitrary images with color control, we produce several photographs on different substrates. Figure 4a exhibits a printed 'The Weeping Woman' on a polished single-crystalline Si wafer. Next, printing on an unpolished backside of 4-inch Si wafer has also been performed, as shown in Fig. 4b. The presented colors are insensitive to surface roughness. More interestingly, owing to diffuse reflection, we find that the generated colors on rough surfaces delivers more uniform brightness than that on polished ones. Our technique also works on flexible substrates. As an example, 50-nm-TiN and 50-nm-TiAlN hybrid films are successively deposited on a 100-μm-thick stainless-steel foil. As shown in Fig. 4c, a famous Chinese calligraphy 'Orchid Pavilion Collection' is printed in a 20×5 $cm^2$ area in 35 seconds. Such flexible foil can be rolled up onto a beverage can, as shown in Fig. 4d. The printing processes are highly reproducible. We fabricate 64 identical samples. The difference among them is nearly not perceptible by the human eye. In practical applications, the durability of structural colors — for example, in terms of resistance to bleaching, abrasion and corrosion— is also vital for any decoration technology. The aging tests for our laser-written structural colors are independently performed through ultraviolet light bleaching, salt fog and high-temperature high-humidity corrosion, and adhesion test (see Methods). After exposing in these extreme environments for 120 h, the adhesion of the structural colors remains to be excellent, and the caused color differences are $\Delta E_{a*b}<5$, which is typically considered an acceptable match in commercial reproduction on printing presses. The plasmonic colors need to be appropriately embedded or protected by coatings, [3] but in our case, the laser-produced oxide film already behaves as a protective coating. Therefore, our technique meets all the requirements of structural colors in practical applications as listed in Fig. 2(g).

In conclusion, we have demonstrated ultrafast laser coloring on optical absorbers that are composed of TiAlN on TiN hybrid film with a total thickness of < 120 nm. Using the conventional nanofabrication techniques to generate colorful patterns on thin-film absorbers generally requires multiple steps of contact photolithography with alignment. [45] However, a single step is sufficient printing colors by ultrafast lasers. Laser-induced oxidation on the surface of absorbing dielectric films alters the reflectance spectra of the TiAlN-TiN absorbers and thus resulting in oxide and remained TiAlN thickness-dependent colors. The formation of a transparent oxide layer significantly widens the gamut of the hybrid thin films. Meanwhile, the extremely hard ceramic material TiAlN exhibits excellent resistance in mechanical, thermal, chemical and abrasion. As a result, the laser printed colors on TiAlN-TiN films present wide gamut, wide viewing angles, high-throughput, high-resolution and high-durability, rendering them to be appealing in practical applications.

**Acknowledgements**

This research was supported by the National Key Research and Development Program of China (2017YFA0205700), the National Natural Science Foundation of China (NO. 61927820, No.12004314, No.62105269). Jiao Geng is supported by Zhejiang Province Selected Funding for Postdoctoral Research Projects (No. ZJ2021044), and China Postdoctoral Science Foundation (2021M702916). Liping Shi is supported by the open project program of Wuhan National Laboratory for optoelectronics No. 2020WNLOKF004 and Zhejiang Provincial Natural Science Foundation of China under Grant No. Q21A040010. The authors thank the technical support from Center for Micro/Nano Fabrication, from instrumentation and Service Center for Physical Sciences and from Instrumentation and Service Center for Molecular Sciences at Westlake University. We thank Mr. Danyang Zhu and Miss. Shan Wu for taking the photographs of the fabricated samples. We thank Miss. Yingchun Wu for milling the samples by focused ion beam. We thank Miss. Ruiqian Meng and Prof. Dianyi Liu for assistance of measuring the spectra.

## 4 Methods

### 4.1 Experimental setups.

In our experiments, the surface coloring is carried out by a home-built laser marking machine. A 1030 nm picosecond laser (Amplitude) with tunable repetition rate as well as pulse energy was focused by a lens (focal length = 20 cm) onto the samples. The laser spot diameter was 120 $\mu$m. The samples were exposed in ambient environment. The laser beam was rasterly scanned by two galvo mirrors for patterning via surface oxidation. The laser polarization was rotated by a quarter-wave plate to be circular, which is in order to avoid the formation of periodic ripples.

### 4.2 Sample fabrication and characterization.

The TiN films were deposited on silicon wafers by RF magnetron reactive sputtering at 300◦C, power of 600 W, and N2 flow of 14 sccm and Ar of 56 sccm. The TiAlN films were deposited as the same parameters for TiN while the power of Al target is DC 300 W. The AlN films were coated as the same for TiAlN while the Ti target was off. The thickness of our films was measured with a profiler (Stylus). According to the measured thickness, the permitivities were retrieved with a variable angle spectroscopic ellipsometer (Woollam). The scanning electron images, and energy dispersive x-ray spectra were measured by a field-emission scanning electron microscope (FE-SEM, Carl Zeiss, Gemini450). The x-ray diffraction analysis of the TiAlN-on-TiN coating on silicon wafer was carried out by Bruker D8 Discover.

### 4.3 Simulation and measurement of reflectance spectra and calculation of chromaticity.

The numerical computations were performed by using commercial software finite-difference-time-domain (FDTD) method (Lumerical FDTD solutions software package). The light source was a depolarized plane wave spanning from 400 nm to 2 µm. The top and bottom boundaries were perfectly matched layers (PML), while periodic boundary conditions were applied in x-y plane. One monitor was set 2 µm above the source to acquire the reflection spectra. The permitivities of the materials involved were measured with the aforementioned ellipsometer. The measured reflectance spectra at near-normal incidence were recorded using a Shimadzu UV-VIS-IR spectrophotometer (UV3600Plus+UV2700). The angle-dependent reflectance and transmittance spectra were measured by an Agilent Cary spectrophotometer (Cary 6000I, UV-Vis-NIR System). The colors in x y chromaticity diagram were calculated based on the simulated and measured reflectance spectra and the colour matching functions defined by the CIE 1931. [49] The spectral power distribution can be expressed as: $P(\lambda) = I(\lambda)R(\lambda)$, in which $I(\lambda)$ is the radiance spectrum of light source and $R(\lambda)$ is the reflectance spectrum.

### 4.4 color fastness tests

The color fastness tests include salt fog test, double-85 test, light bleaching test and adhesion test. The salt fog test is performed according to the standard of GB/T 10125-2021 for 120 hours. The density of NaCl is 50 g/L with a sedimentation rate of 2 ml/(80cm2h). The double-85 test is carried out in the standard of GB/T 1740-2007 for 120 hours. The temperature and humidity are 85℃ and 85%. The light bleaching test (GB/T 1865-2009) is irradiated by 340 nm UV light with an irradiance of 0.51 W/(m2nm) for 120 hours. The temperature and humidity are 65 ◦C and 50% RH, respectively. The adhesion test is performed by using 3M tapes according to the standard of GB/T 9286-2021. The color differences ($\Delta Ea*b$) between the pristine and tested samples are measured by an integrating sphere spectrophotometer using the D65 light source.

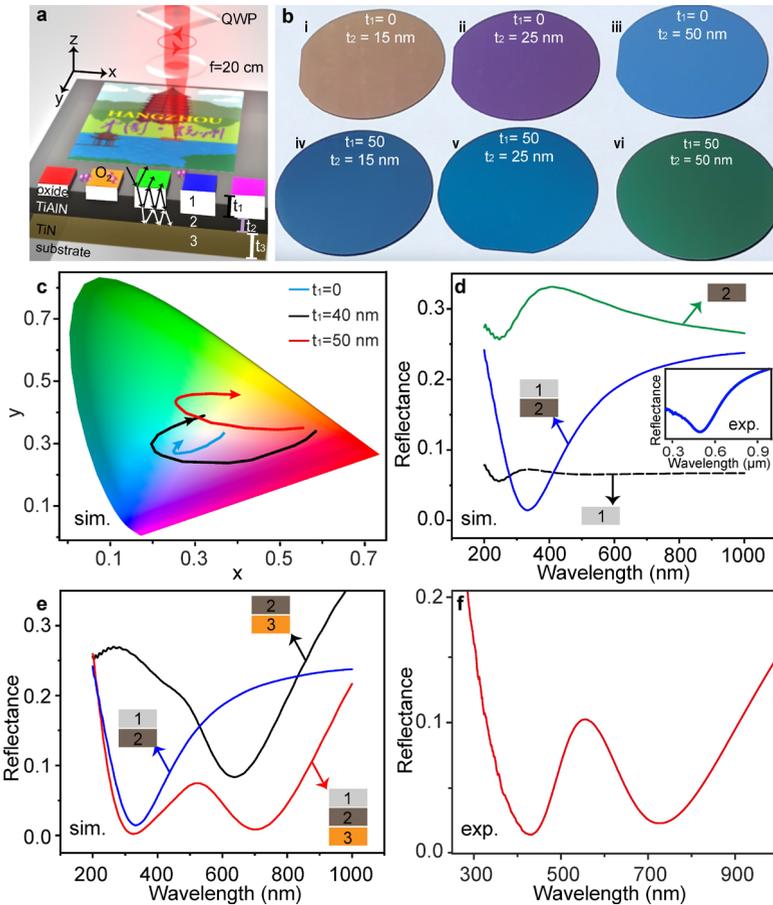

**Fig. 1 Scheme of surface coloring on ultrathin films by ultrashort lasers.** The structural colors originate from laser-induced oxidation on highly absorbing while extremely hard ceramic TiAlN film. The interference of light that transmits and reflects at the interfaces gives rise to strongly resonant absorption and displays tunable colors. The TiN film behaves as a reflector. $t_1$: thickness of laser-induced oxide layer (medium 1); $t_2$: thickness of remained TiAlN film (medium 2); $t_3$: thickness of TiN (medium 3). QWP: quarter-wave plate. **b**, Photograph of surface coloring with various TiAlN and AlN coatings on $t_3$=50 nm TiN. The substrates are Si wafers or glasses. **c**, Numerically simulated CIE 1931 color coordinates for variable $t_2$ (TiAlN) and several different thickness of $Al_2O_3$ coating ($t_1$ = 0, 40, 50 nm). $t_2$ gradually increases from 10 nm to 50 nm with a step of 2 nm along the arrows. **d**, Simulation of normal-incidence reflectance spectra of $Al_2O_3$, TiAlN, and their stack coating on a glass substrate, respectively. The thickness of each coating is 40 nm. Inset: measured reflectance spectrum of 40 nm AlN coating on 40 nm TiAlN on a glass substrate. **e**, simulated reflectance spectra at normal incidence of $Al_2O_3$-on-TiAlN (blue curve), TiAlN-on-TiN (black curve), and $Al_2O_3$-on-TiAlN-on-TiN (red curve), respectively. The thickness of $Al_2O_3$ (medium 1), TiAlN (medium 2) and TiN (medium 3) are 40nm, 40nm, and 50 nm, respectively. **f**, measured reflection spectrum of the sample shown in **b(vi)**. Please note that, in experiments the lossless dielectric coating (medium 1) **b, d, f** is AlN, because it is more convenient to use the same targets as that for TiAlN during the sputtering process, while in simulations, **c-e**, medium 1 is set as $Al_2O_3$ to match the laser-produced oxide layer on TiAlN.

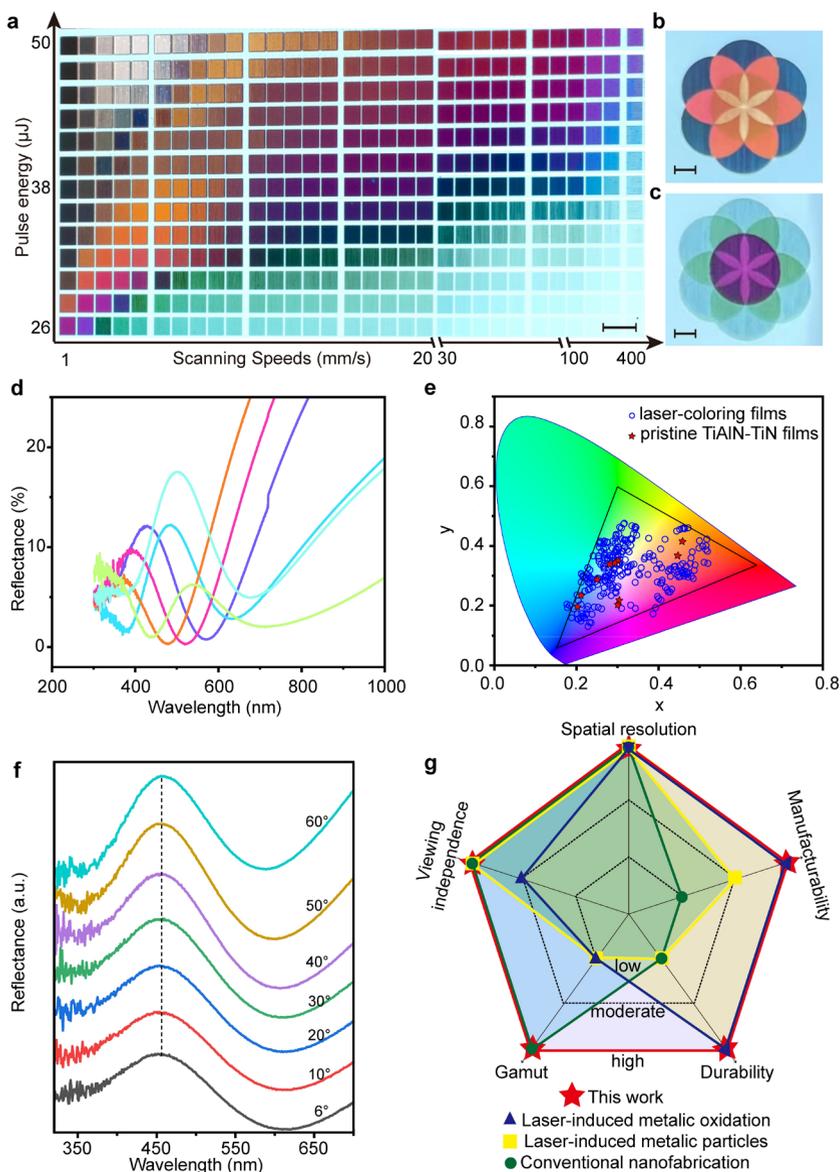

**Fig. 2 Optical properties of laser-printed colors on TiAlN-TiN hybrid films.** **a**, matrix palette produced with various pulse energies and scanning speeds. The thickness of TiAlN and TiN are 60 nm and 50 nm, respectively. **b, c**, seven circles that are independently written. The colors in the overlapping areas are changed. The scale bars are 5 mm. **d**, near-normal incidence (6°) reflectance spectra of several representative laser-written colors. **e**, CIE *x-y* chromaticity diagram, comparing the colors of pristine bilayer TiAlN-TiN films (stars) and laser-produced colors (circles). The thickness of TiAlN coatings ranges from 15 nm (in the yellow zone) to 60 nm (in the green zone) with a step size of 5 nm. **f**, reflectance spectra versus incident angles. **g**, technology-performance indicators expressing on low, moderate and high level that provides an overview of the conventional nanofabrication and state-of-the-art laser coloring. [3, 37]

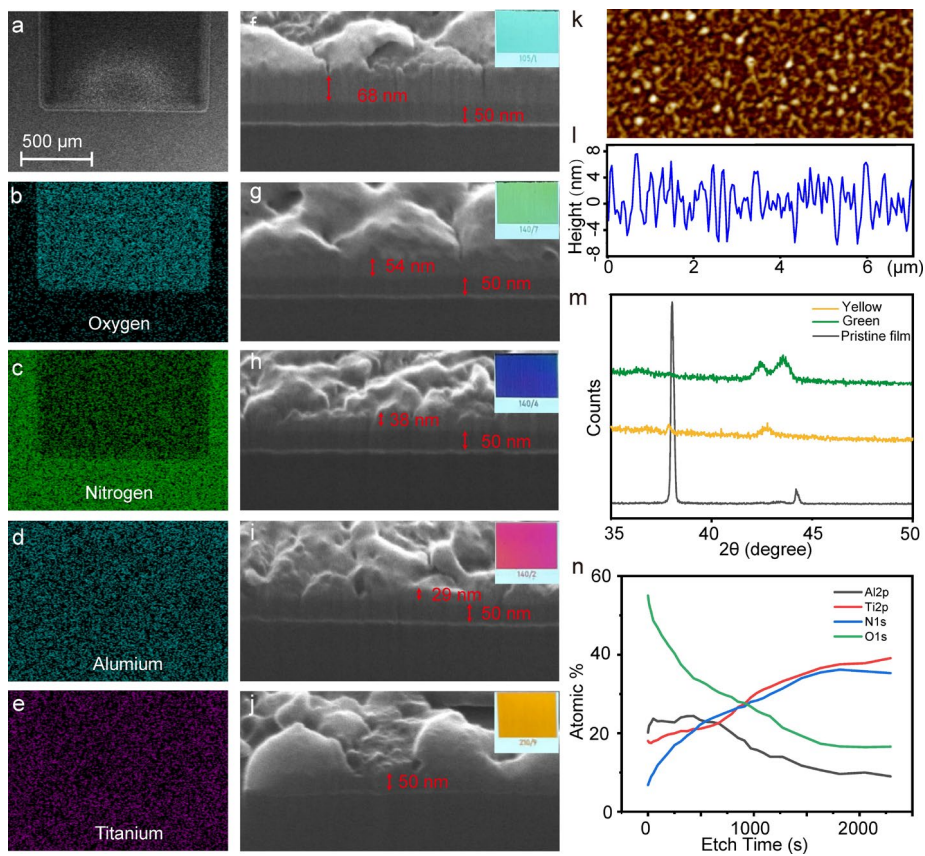

**Fig. 3 Material properties of laser-printed colors.**, **a-e**, SEM image and corresponding EDX maps of a laser-written area. **f-j**, cross-sectional view of films that are obtained by different scanning speeds or pulse energies. The pulse energy and scanning speed from **f** to **j** are 21, 28, 28, 28, 42 µJ, and 1, 7, 4, 2, 9 mm/s, respectively. The insets show their corresponding colors. **k, l**, surface topography analysis by AFM. **m**, XRD spectra, and **n**, depth profile XPS analysis of oxygen distribution.

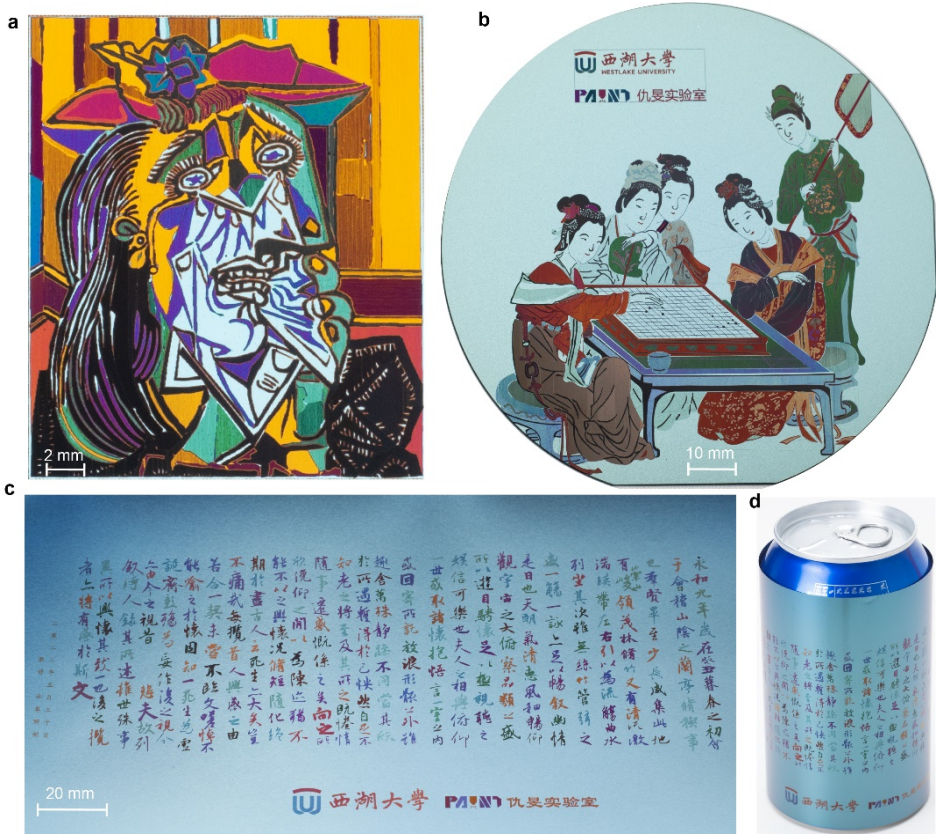

**Fig. 4 Laser coloring of large areas for painting**, **a-c**, Photographs of structural color- based patterns which are printed on different substrates, including polished **a**, unpolished **b** 4-inch Si wafers and large-scale stainless-steel foil **c**. The thickness of TiAlN and TiN are 60 nm and 50 nm, respectively. **d**, the same photograph in **c** while being rolled on a drink can.